\documentclass[aip, jap, reprint, floatfix]{revtex4-1}

\usepackage{graphicx}
\usepackage{amssymb}
\usepackage{amsmath}
\usepackage{comment}
\usepackage{url}
\usepackage[usenames, dvipsnames, svgnames, table]{xcolor}
\usepackage{siunitx}

\newcommand{\field}{B}
\newcommand{\tracewidth}{w}
\newcommand{\tracelength}{\ell}
\newcommand{\distance}{d}
\newcommand{\fluxquantum}{\Phi_0}
\newcommand{\vortexnumber}{N}

\newcommand{\threshold}{\mathrm{th}}
\newcommand{\critical}{\mathrm{c}}
\newcommand{\ambient}{\mathrm{a}}
\newcommand{\adr}{\mathrm{ADR}}
\newcommand{\magnetarray}{\mathrm{m}}
\newcommand{\internal}{\mathrm{i}}
\newcommand{\coupling}{\mathrm{c}}
\newcommand{\resonator}{\mathrm{r}}

\newcommand{\columbia}{Department of Physics, Columbia University, New York, NY 10027, USA}
\newcommand{\starcryo}{STAR Cryoelectronics, Santa Fe, NM 87508, USA}
\newcommand{\asusese}{School of Earth and Space Exploration, Arizona State University, Tempe, AZ 85287, USA}
\newcommand{\asuphysics}{Department of Physics, Arizona State University, Tempe, AZ 85287, USA}
\newcommand{\cardiff}{School of Physics and Astronomy, Cardiff University, Cardiff, Wales CF24 3AA, UK}
\newcommand{\jpl}{Jet Propulsion Laboratory, Pasadena, CA 91109, USA}
\newcommand{\caltech}{Division of Physics, Mathematics, and Astronomy, California Institute of Technology, Pasadena, CA 91125, USA}
\newcommand{\usc}{Department of Physics and Astronomy, University of Southern California, Los Angeles, CA 90089, USA}

\begin{document}

\title{Magnetic field dependence of the internal quality factor and noise performance of lumped-element kinetic inductance detectors}

\author{D.~Flanigan}
\email{daniel.flanigan@columbia.edu}
\affiliation{\columbia}

\author{B.~R.~Johnson}
\affiliation{\columbia}

\author{M.~H.~Abitbol}
\affiliation{\columbia}

\author{S.~Bryan}
\affiliation{\asusese}

\author{R.~Cantor}
\affiliation{\starcryo}

\author{P.~Day}
\affiliation{\jpl}

\author{G.~Jones}
\affiliation{\columbia}

\author{P.~Mauskopf}
\affiliation{\asusese}
\affiliation{\asuphysics}
\affiliation{\cardiff}

\author{H.~McCarrick}
\affiliation{\columbia}

\author{A.~Miller}
\affiliation{\usc}

\author{J.~Zmuidzinas}
\affiliation{\jpl}
\affiliation{\caltech}

\date{\today}

\begin{abstract}
We present a technique for increasing the internal quality factor of kinetic inductance detectors (KIDs) by nulling ambient magnetic fields with a properly applied magnetic field.
The KIDs used in this study are made from thin-film aluminum, they are mounted inside a light-tight package made from bulk aluminum, and they are operated near 150~mK.
Since the thin-film aluminum has a slightly elevated critical temperature ($T_\critical = \SI{1.4}{K}$), it therefore transitions before the package ($T_\critical = \SI{1.2}{K}$), which also serves as a magnetic shield.
On cooldown, ambient magnetic fields as small as approximately \SI{30}{\micro T} can produce vortices in the thin-film aluminum as it transitions because the bulk aluminum package has not yet transitioned and therefore is not yet shielding.
These vortices become trapped inside the aluminum package below \SI{1.2}{K} and ultimately produce low internal quality factors in the thin-film superconducting resonators.
We show that by controlling the strength of the magnetic field present when the thin film transitions, we can control the internal quality factor of the resonators.
We also compare the noise performance with and without vortices present, and find no evidence for excess noise beyond the increase in amplifier noise, which is expected with increasing loss.
\end{abstract}

\maketitle


Kinetic inductance detectors\cite{Day2003} (KIDs) are superconducting resonators designed to detect photons.
In the planar lumped-element resonator geometry\cite{Doyle2010} used in this study, the resonance is produced by an interdigitated capacitor and a meandered inductor that is also the radiation absorber.
The devices are made from thin-film aluminum.
The suitability of KIDs as a detector technology for photometry depends in part on the fact that they can exhibit high resonator quality factors $Q_\resonator$.
By tuning each resonator to a unique frequency and taking advantage of the narrow bandwidth corresponding to high $Q_\resonator$, hundreds or thousands of KIDs may be read out on the same feed line using frequency division multiplexing.
The resonator quality factor is related to the internal loss of the resonator:
$Q_\resonator^{-1} = Q_\internal^{-1} + Q_\coupling^{-1}$,
where the inverse of the internal quality factor $Q_\internal$ is the sum of all internal loss contributions, and the coupling quality factor $Q_\coupling$ is inversely related to the strength of the coupling to the feed line.
To maintain excellent noise performance and multiplexing capability, it is desirable for the internal loss to be dominated by quasiparticles and not by other sources.\cite{Zmuidzinas2012}

\begin{figure*}[t]
\centering
\includegraphics[width=\textwidth]{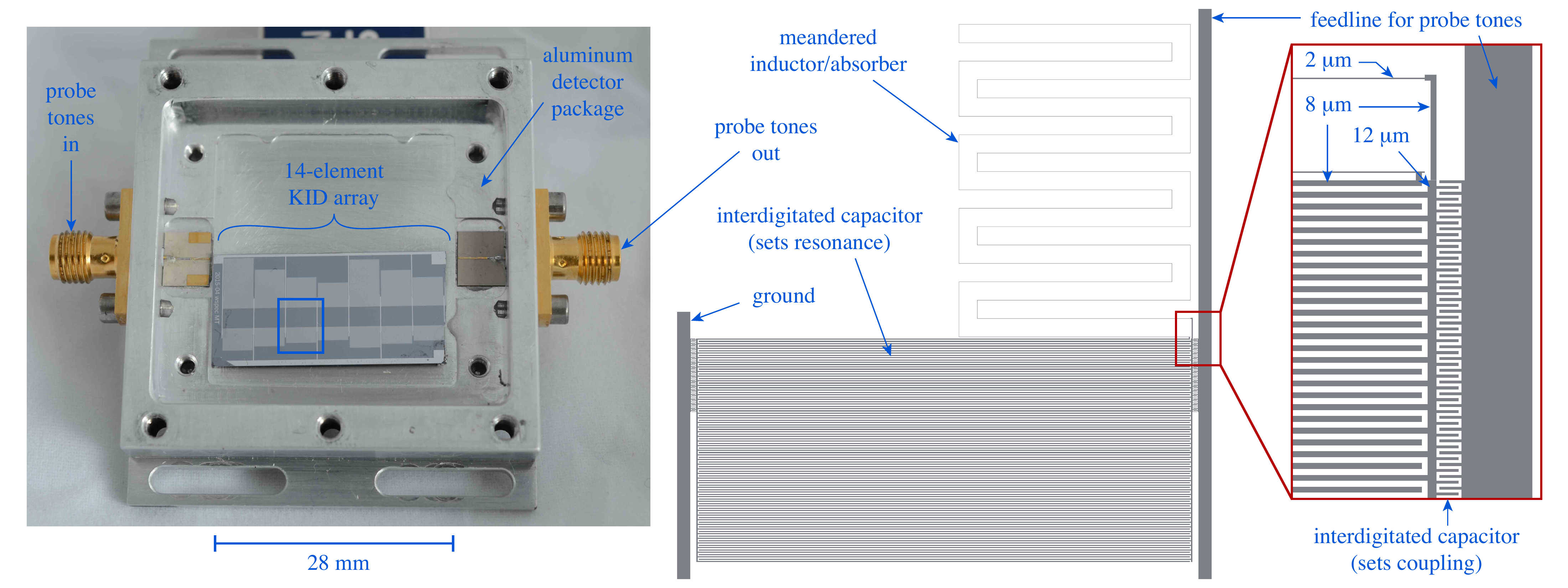}
\caption{
\textbf{Left:}
A photograph of the detector module tested in this study.
The package lid is removed so the KID array is visible.
Metal clips, also not shown here, are used to hold the KID array in place.
\textbf{Center:}
A scale drawing of the lumped-element KID in the blue box on the left.
\textbf{Right:}
Detail of the center panel, showing all of the trace widths used in the resonator.
Given the strength of the ambient magnetic field, unwanted vortices can form in the \SI{12}{\micro m} trace yielding low $Q_\internal$. 
}
\label{fig:experiment}
\end{figure*}

With some device architectures we have observed internal quality factors that are significantly lower than expected.
Common loss mechanisms, such as quasiparticles and two-level systems on the dielectric surfaces, were experimentally ruled out for these devices, so we started considering effects from ambient magnetic fields, such as the Earth's magnetic field.
Initially, this seemed unlikely because our KIDs are operated in a package made from bulk aluminum.
As a type-I superconductor, aluminum expels external magnetic fields making it an excellent magnetic shield.
Indeed, after the system is cooled well below the bulk aluminum critical temperature of \SI{1.2}{K} the KIDs do not detectably respond to externally applied magnetic fields.
This includes the aforementioned low-$Q_\internal$ resonators.
Therefore, we cultivated the following hypothesis.
Thin films of type-I superconductors permit flux entry in the form of vortices.\cite{Tinkham1963,Dolan1974}
These vortices produce loss in thin-film superconducting resonators\cite{Song2009,Wang2009,Mazin2010,Bothner2011,Bothner2012,deGraaf2012,Nsanzineza2014,Chiaro2016} because an alternating current in a thin-film trace produces an oscillating Lorentz force on a vortex and the vortex motion is dissipative.\cite{Song2009}
Since the thin film used in the KID has a critical temperature $T_\critical = \SI{1.4}{K}$, and thus transitions before the package when the system is cooled, vortices formed in the momentarily un-shielded film become trapped inside the aluminum package and persist when the system is cooled far below $T_\critical$.
Thus, the presence of vortices at the KID operating temperature ($\sim \SI{150}{mK}$) depends on the field present when the aluminum film transitions.
We tested this hypothesis by varying the strength of the ambient field at \SI{1.4}{K} and measuring $Q_\internal$ at the KID operating temperature.
We manipulated the unavoidable ambient magnetic field in the room using an array of permanent magnets with opposite polarity thereby creating a movable magnetic field null.
The experimental setup and the results are described below.


The resonators used in this study are lumped-element kinetic inductance detectors (LEKIDs) lithographically patterned from a \SI{20}{nm} thick aluminum film on a \SI{500}{\micro m} thick high-resistivity, float-zone silicon substrate.
They were designed for astrophysical measurements at millimeter wavelengths.
The detectors tested in this study were not optically illuminated and were instead mounted inside a light-tight aluminum package with copper tape covering the seam to prevent light leaks.
The package was machined from QC-10, an aluminum alloy for which we have measured a critical temperature near that of bulk aluminum (\SI{1.2}{K}).
The left panel of Figure~\ref{fig:experiment} is a photograph of the KID array in the package.
Fourteen resonators were patterned in this array.
For this study we focused on just three of these resonators with resonance frequencies $f_\resonator$ of 78, 116, and \SI{161}{MHz}.
The center and right panels of Figure~\ref{fig:experiment} are drawings of one resonator that show the various trace widths used in different regions.

Inside the cryostat, the package was mounted to a gold-plated copper plate that is thermally connected to the cold stage of an adiabatic demagnetization refrigerator (ADR) backed by a helium pulse tube cooler.
The shells of the cryostat are aluminum, and, except where noted below, none of the materials near the package are ferromagnetic or high-permeability.

The ambient field of the room in the region of the package was measured to be downward to within \SI{10}{\degree} of vertical.
We do not consider any effect of the in-plane component of the magnetic field and refer hereafter to only the vertical component of the field, which is normal to the aluminum film.
All reported fields were measured using a gaussmeter that uses a calibrated single-axis Hall probe.\cite{gaussmeter}
Taking the upward direction to be positive, the ambient field is $\field_\ambient = -30 \pm 1 \, \si{\micro T}$.
While collecting data over several weeks, we observed daily drifts of only a few \si{\micro T} in measurements of the ambient field near the cryostat, so it is stable in time during a given measurement.

We created a magnetic field normal to the KIDs using an array of seven small NdFeB permanent magnets mounted outside the cryostat.
The magnets were arranged in a triangular lattice \SI{70}{mm} in diameter to make the magnetic field and the desired null uniform and large relative to the array size, which is \SI{28}{mm} by \SI{13}{mm}.
Any lateral variation in the field is 10\% or less. 
The field produced by this lattice $\field_\magnetarray(\distance)$ was measured as a function of distance $\distance$ from the center plane of the magnets.
Over the relevant range of distances between the magnets and the KID array, we measured $\field_\magnetarray \sim \distance^{-3}$ as expected.

To record a data set, we first establish a magnetic field configuration by positioning the magnet array some chosen distance from the KID array.
Because the cryostat shells are made from aluminum, the ambient magnetic field and the magnetic field from the permanent magnet lattice should enter the cryostat unaltered.
The total magnetic field that we report is
$\field = \field_\ambient + \field_\magnetarray$.
Note that $\field$ can be zero if the created null is positioned right at the KID array.
After setting the field, we cycle the ADR, let the package and KID array cool well below $T_\critical$, regulate the temperature of the package at $153 \pm \SI{4}{mK}$, then collect data.
For each resonator we first, using a ROACH-based readout, sweep the readout tone frequency across the resonance and fit the data to a resonator model,\cite{McCarrick2014} then set the readout tone to the resonance frequency obtained from the fit and collect time-ordered data for \SI{30}{s}.
The data set ultimately yields a value for $Q_\internal$ and a noise spectrum for that magnetic field configuration.
All measurements were recorded using a constant readout tone power of approximately \SI{-100}{dBm} on the feedline, below the onset of nonlinear effects in the resonators.
This process was repeated for a range of distances.
For comparison, we also recorded data with a five-sided mu-metal shield surrounding the cryostat.
The contribution of the ambient field to the interior of the mu-metal shield was measured to be less than \SI{1}{\micro T}, and we take it to be zero when the shield is used.


\begin{figure}[t]
\includegraphics[width=\columnwidth]{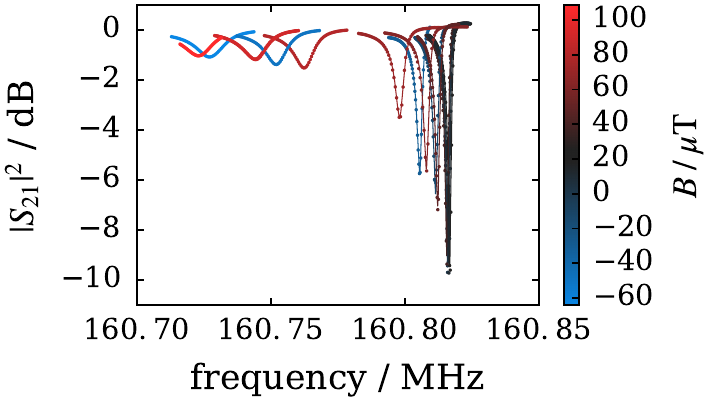}
\caption{
The points are forward scattering parameter $S_{21}$ data from sweeps of the readout tone across the \SI{161}{MHz} resonance.
The data have been normalized to 1 off-resonance using the fits to the resonator model, which are plotted as lines.
The color bar shows the field ($\field_\ambient + \field_\magnetarray$) in which the resonator was cooled.
}
\label{fig:s21_vs_f}
\end{figure}

\begin{figure}[b]
\includegraphics[width=\columnwidth]{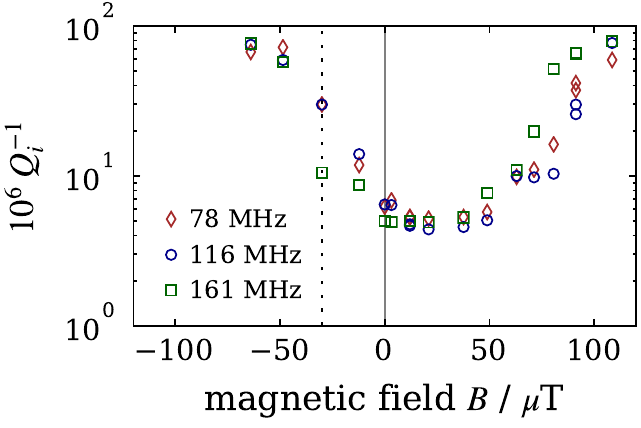}
\caption{
The inverse internal quality factor versus magnetic field ($\field_\ambient + \field_\magnetarray$), plotted for three resonators.
The vertical gray line marks the field condition when a mu-metal magnetic shield is placed around the cryostat and no magnet array is applied.
The dotted black line marks the field condition with no magnetic shield present and no magnet array.
The points to the left of the dotted black line were recorded with the magnet array polarity reversed so that it augmented the ambient field.
The minimum is likely shifted away from zero because the Heli-Coil inserts in the cold plate of the cryostat can produce a $\sim \SI{25}{\micro T}$ field.
}
\label{fig:iQi_vs_B}
\end{figure}

Figure~\ref{fig:s21_vs_f} shows the behavior of one resonator as $\field_\magnetarray$ is varied.
At higher field magnitudes, $f_\resonator$ and $Q_\internal$ both decrease, while $Q_\coupling$ does not change.
As shown in Figure~\ref{fig:iQi_vs_B}, the loss minimum for all three resonators occurs over a range of fields centered near $\field = 0$, and the loss increases as the field magnitude departs from this central value.
This result is consistent with previous studies of vortices in thin films, which have generally found that increasing field magnitude creates both higher vortex density in narrow strips and higher loss in resonators.
Direct imaging of the field near narrow strips of thin-film niobium\cite{Stan2004} and YBCO\cite{Kuit2008} has shown that few or no vortices enter the strip below a threshold field magnitude $\field_\threshold$, which varies with the trace width $\tracewidth$ approximately as
$\field_\threshold \sim \fluxquantum \tracewidth^{-2}$,
where $\fluxquantum$ is the flux quantum.
Measurements of the vortex-induced loss in aluminum and rhenium thin-film resonators cooled in a magnetic field normal to the film showed that the field had no effect on the loss below a threshold magnitude, and that well above this level the loss was approximately proportional to the excess magnitude above the threshold.\cite{Song2009}
The entry of even a single vortex into a region of high current flow in a resonator can cause significant loss.\cite{Nsanzineza2014}

In Figure~\ref{fig:iQi_vs_B}, the center of the low-loss region is offset from zero by about \SI{25}{\micro T}.
We believe that this offset is caused by fields not included in the calculation of $\field$.
First, during the course of these measurements we discovered that the stainless steel Heli-Coil inserts in the millikelvin stage plate are magnetized.
While this Heli-Coil field is not constant across the KID array, its magnitude and direction approximately account for the observed offset.
Second, while the ADR is well-shielded with Vanadium Permendur, field leakage is possible and expected.
To estimate how much our result is influenced by stray fields from the ADR, we conducted a separate measurement of the vertical component of the field ($\field_\ambient + \field_\adr$).
Because our Hall probe cannot operate at cryogenic temperatures we made the field measurement \SI{6}{cm} below the package just outside the cryostat.
At peak current, when the package is at \SI{3}{K}, a detectable $\field_\adr$ is produced by the ADR coil.
However, $\field_\adr$ decreases during the ADR cycle because the current in the coil decreases.
The measured field ($\field_\ambient + \field_\adr$) returns to within the measurement uncertainty of $\field_\ambient$ when the package reaches \SI{1.4}{K}, indicating that $\field_\adr$ is small at the relevant point in the cycle.
Our conclusion is that the ADR field could produce a shift in the center of the low-loss region shown in Figure~\ref{fig:iQi_vs_B}, but it is likely to be a smaller source of systematic error than field from the Heli-Coil inserts.

Interpreting the offset in this way, the threshold field for vortex entry is $\field_\threshold \approx \SI{30}{\micro T}$.
As shown in Figure~\ref{fig:experiment}, the widest traces in these resonators are \SI{12}{\micro m}; these are located only where the coupling capacitor runs along part of the much larger capacitor that sets the resonance frequency.
The threshold field for this width is expected to be $\fluxquantum \tracewidth^{-2} = \SI{14}{\micro T}$ (up to a factor that is of order unity).
Since these wider traces are near the junction with the inductor, most of the current will flow through them on the way into the \SI{8}{\micro m} wide interdigitated capacitor tines, so we expect vortex entry here to produce loss.
Previous measurements of similar lumped-element resonators with a maximum trace width of \SI{8}{\micro m} consistently exhibited high $Q_\internal$.\cite{McCarrick2014}
The crucial difference seems to be that the \SI{12}{\micro m} trace here permits vortex entry at a threshold field less than the ambient field, while the \SI{8}{\micro m} traces did not.

\begin{figure}[t]
\includegraphics[width=\columnwidth]{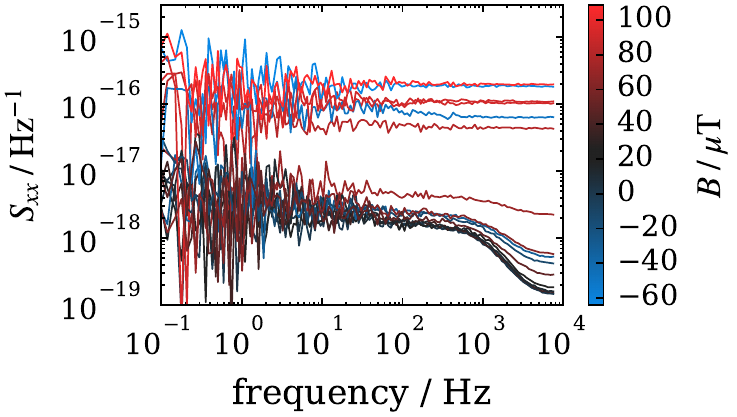}
\caption{
The spectral density of the fractional frequency time series data from the \SI{161}{MHz} resonator.
The color scale corresponds to the magnetic field ($\field_\ambient + \field_\magnetarray$).
To more clearly show the trend at low frequencies, the lowest 15 harmonics of the \SI{1.412}{Hz} pulse tube cooler frequency have been masked in all of the spectra.
The color bar is the same as Figure~\ref{fig:s21_vs_f}.
}
\label{fig:Sxx_vs_f}
\end{figure}

In SQUIDs, the presence of vortices is known to produce flux noise with a typical $1 / f$ spectral density.\cite{Dantsker1997}
To investigate the possibility of vortices producing excess noise in the resonators, we decomposed the on-resonance time-ordered data into two real time series corresponding to the fractional frequency shift $x(t)$ and inverse internal quality factor $Q_\internal^{-1}(t)$.
The spectral density $S_{xx}(f)$ of the $x(t)$ data is shown in Figure~\ref{fig:Sxx_vs_f}.
Larger field magnitudes correspond to higher loss, and thus a higher amplifier noise level.
Besides this expected effect of lower $Q_\internal$, we see no evidence for additional contributions to the noise due to the presence of vortices.
Only amplifier noise is visible in the internal loss fluctuation spectra (not shown here).

To verify that the superconducting detector package is an effective magnetic shield, we altered the magnetic field after the package had fully cooled and looked for changes in $Q_\internal$ and $f_\resonator$.
We cooled the package in an initial field condition near the center of the low-loss region in Figure~\ref{fig:iQi_vs_B}, collected the nominal data set, moved the magnet array to establish a new high-field condition at the package, and then collected a second data set.
Between these data sets, neither $Q_\internal$ nor $f_\resonator$ changed significantly indicating the perturbation in the applied magnetic field condition did not affect the resonators either through vortex entry or kinetic inductance non-linearity.\cite{Healey2008, Zmuidzinas2012}
Note that these second points are not shown in Figure~\ref{fig:iQi_vs_B}.
The observation that some vortices remain in the film even when the package is shielding the resonators is consistent with the hysteretic magnetization curves observed in field-cooled type-I thin films\cite{Chang1963,Miller1964} and with hysteretic loss observed in niobium thin-film resonators.\cite{Bothner2012}
We can use a result of \citet{Stan2004} to estimate the number of vortices $\vortexnumber$ present just below $T_\critical$ in a trace of width $\tracewidth = \SI{12}{\micro m}$ and length $\tracelength = \SI{1000}{\micro m}$ (this length varies substantially between resonators):
$\vortexnumber
  \approx
  (\field - \field_\threshold) \tracewidth \tracelength / \fluxquantum
  \sim
  300$
at the highest field magnitudes.


In conclusion, several findings presented here should be useful to groups developing KIDs.
First, it is possible to maximize the resonator $Q_\internal$ and therefore the sensitivity of KIDs by nulling ambient magnetic fields with a properly applied magnetic field.
Second, fabricating the detector package from a type-I superconductor with a $T_\critical$ that is higher than that of the film might prevent vortex entry.
Third, a mu-metal magnetic shield surrounding the cryostat is very effective.
Fourth, hardware elements such as Heli-Coil inserts or nickel-plated connectors, which are commonly used near the detector package inside the cryostat, could produce magnetic fields strong enough to yield vortices.\cite{Wang2009,Chiaro2016}
As alternate solutions to the problem, we are also (i) exploring modifications to the devices themselves, such as flux-trapping holes\cite{Bothner2011,Chiaro2016} and fractal geometry,\cite{deGraaf2012} (ii) considering using a Helmholtz coil instead of permanent magnets\cite{Song2009,Mazin2010,Bothner2012,Nsanzineza2014} and (iii) investigating ways of momentarily heating the KID arrays inside the superconducting package to the point where the aluminum film becomes normal and the vortices dissipate.


R.C. is both an author and the owner of STAR Cryoelectronics, where the devices used in this study were fabricated.
H.M. is supported by a NASA Earth and Space Sciences Fellowship.
This research is supported, in part, by a grant from the Research Initiatives for Science and Engineering program at Columbia University to B.R.J.
We thank the Xilinx University Program for their donation of FPGA hardware and software tools used in the readout system. 

\bibliography{references}

\end{document}